# Chiral d-density wave ordered pseudo gap phase of $Bi_2Sr_2CaCu_2O_{8+\delta}$ bilayer. Spin texture and Dzyaloshinskii–Moriya interaction

*Kakoli Bera, U.P.Tyagi, and Partha Goswami(Retd.)\**

*Deshbandhu College, University of Delhi, Kalkaji, New Delhi-110019,India*

*\*E-mail of the corresponding author: physicsgoswami@gmail.com; E-mail of the first author:kakolibera@gmail.com; E-mail of the second author:uptyagi@yahoo.co.in*

**ABSTRACT**

We model $Bi_2Sr_2CaCu_2O_{8+\delta}$(Bi2212) bilayer system by an inversion symmetry broken and time reversal non-invariant Bloch Hamiltonian H. The pseudo-gap phase of the bilayer is assumed to be chiral d-density wave ordered. The focal point of the paper is the theoretical study of the spin-momentum locking (SML), due to the presence of a strong spin-orbit coupling in Bi2212, reported earlier in this phase in a spin- and angle-resolved photoemission spectroscopic measurement. The non-trivial spin texture in **k**-space is found tunable by electric field(and also by intercalation).The Dzyaloshinskii–Moriya interaction (DMI) coefficient, where DMI being important energy for chiral textures like magnetic skyrmions, are calculated to set the stage for detailed studies in a sequel. We also find that(in the nodal region) in the presence of Zeeman field B , as in a s-wave superconductor, for $B \geq B_c$ (*a critical value*) a vortex in this system becomes a non-Abelian anyon with a Majorana zero mode required for the fault tolerant quantum computation.

## 1.Introduction

We investigate the band structure of $Bi_2Sr_2CaCu_2O_{8+\delta}$(Bi2212) bilayer system starting with a time reversal symmetry, and inversion symmetry broken Bloch Hamiltonian involving a term accounting for the effect of coupling between different $CuO_2$ planes. The coupling possesses a very different form in a single layer cuprate such as LSCO or NCCO than when a $CuO_2$ bilayer is present as in Bi2212. The Hamiltonian also involves Rashba spin-orbit coupling (RSOC). The origin of RSOC lies in the fact that whereas one Cu-O layer has Ca ions above and, Bi-O ions below, in the unit cell of the other layer this situation is reversed. This leads to a nonzero electric field within the unit cell. A recent spin- and angle-resolved photoemission spectroscopy measurement for $Bi_2Sr_2CaCu_2O_{8+\delta}$ had revealed **[1]** a non-trivial spin texture which corresponds to a well-defined direction for each electron real spin depending on its momentum. This spin-momentum locking (SML)suggests the presence of a strong spin-orbit coupling in Bi2212 like a topological insulator. The band structure analysis of this ambivalent system carried out in the paper, corresponding to the bonding and the anti-bonding cases, yields the following: The spin-momentum locking (SML) is evident, While the nodal region of the momentum space can give rise to a spin-down hole current, the anti-nodal region can give rise to spin-up electron current. We find that the common ground between two types of SMLs, corresponding to the bonding and the anti-bonding cases, is that the states of opposite spin are to be found in different parts of the Brillouin zone.

The Dzyaloshinskii–Moriya interaction (DMI) **[1a-c]** is considered as one of the most important energies for specific chiral textures such as magnetic skyrmions. The interaction was introduced first by Stevens[1a] where it was derived as a consequence of the inclusion of spin-orbit coupling in the Heisenberg model. In order to explain the weak ferromagnetic(FM) moments in largely anti-FM $\alpha$-$Fe_2O_3$ crystals, the DMI was later introduced by Dzyaloshinskii[1b]. The inclusion of spin-orbit coupling in a consideration of the super-exchange mechanism to provide the first microscopic derivation was reported by Moriya[1c] later. He introduced a symmetric tensor of second order in the spin-orbit coupling and the Moriya vector **$D_{ij}$** which is proportional to the first power of the spin-orbit coupling and an antisymmetric vector. In this communication, we initiate an investigation in similar lines starting with the aforementioned Bloch Hamiltonian for Bi2212 bilayer system. In the presence of a strong spin-orbit coupling there is likely-hood of occurrence of the anomalous quantum Hall effect (AQHE). This is observed in 2-D systems at low temperatures with very strong spin-orbit coupling(no magnetic fields is required) in which the transverse Hall resistance undergoes unexpected quantum transitions accompanied by a considerable drop in longitudinal resistance. The intrinsic AHE can be expressed in terms of the Berry-phase curvatures and it is therefore an intrinsic quantum-mechanical property of a perfect crystal. An extrinsic mechanism, skew scattering from disorder, tends to dominate the AHE in highly conductive ferromagnets. Our main aim in this paper is to perform band structure analysis in detail and reporting preliminaries of spin-texture and Moriya-like investigation [1c] for Bi2212 motivated by the findings in**[1]**. The later one paves the way to explore the possibility of skyrmions in the system. The program of ours is driven not only by the fundamental interest in exotic physical systems, but also by its potential applications to next-generation memory, logic, and neuromorphic computing devices. We investigate the possibility of occurrence of AQHE in Bi2212 with our model Hamiltonian in a sequel to this communication. Furthermore, as in the Dirac system, to realize non-abelian anyons we introduce here the z-direction oriented TRS breaking Zeeman field ($H_z = \sigma g \mu_B B$) in the DSC(d wave superconductor with gap $\Delta^{sc}(k) = \Delta^{sc}_0 \sin(k_x a) \sin(k_y a)$). We find that (in the nodal region), as in a s-wave superconductor, for $B \geq B_c$ (*a critical value*), a vortex in this system becomes a non-Abelian anyon with a Majorana zero mode required for the fault tolerant quantum computation.

The conventional superconductors (SC) involve symmetric s-wave spin-singlet pairing of electrons by phonon-mediated attractive interaction, while the unconventional SCs require a long-range interaction [2] and have lower symmetry Cooper pairs. The Chiral SCs [3] are interesting instances of unconventional SCs where Cooper pairs with finite angular momentum circulate around a unique chiral axis. This leads to spontaneous breaking of time-reversal symmetry (TRS). The widely known member of this class is bulk $Sr_2RuO_4$, conjectured to be p-wave SC displaying spin-triplet pairing ; the chiral singlet SCs were believed to be very few and far between. Perhaps, the strongly correlated heavy fermion SC such as $URu_2Si_2$[4] is a known example. The general belief regarding the featuring of spin-triplet p-wave pairing by the chiral SCs was actually based in the absence of a drop in NMR Knight shift[5], the broken TRS found in muon spin relaxation[6], and Kerr rotation experiments [7]. The NMR experiments [8-10] performed in the recent past, however, indicate utterly compelling irreconcilability with a number of odd-parity (pseudo-)spin triplet order parameters leading to the resurrection of the spin-singlet pairing scenario [11].It must be mentioned that before these issues came to the fore, the nature of the pseudo-gap(PG) phase of the cuprate high temperature superconductors had already posed a number of unexplained puzzles. Below a characteristic temperature $T^*$ but higher than the SC transition temperature $T_c$, the excitation spectra showing a gap was first noticed by the relaxation rate of nuclear magnetic resonance[12] and then by many other transport and spectroscopic measurements [13]. But the most direct observation of this gap structure was shown by the ARPES[14]. The energy gap appears near the anti-nodal region, of the two-dimensional Brillouin zone (BZ) of the cuprate. The ARPES spectra at the anti-nodal region does not have the usual particle-hole symmetry associated with traditional superconductors. This asymmetric anti-nodal gap onsets at $T^*$ and it persists all the way to the SC phase[15,16].There are four disconnected segments of Fermi surface near the nodal region. These segments called Fermi arcs, which are believed to be part of a small pocket [17,18],have been reported to have their length not sensitive to temperature[19]. This presence of finite fraction of Fermi surface is consistent with the Knight shift measurement [20] showing a finite density of states (DOS) after the superconductivity is suppressed. Below $T_c$ the gap at anti-node merges with the SC gap. With the aim to explain the nature of PG, many reports have appeared in not so distant past, where charge density waves (CDW) or spin density waves (SDW) were related to SC and PG phases[21-28]. On another level, the pseudo-gap is a distinct phase akin to an unconventional metal or, a symmetry preserved/broken state [29-41]. The onset of the pseudo-gap is defined by the opening of an anti-nodal gap and the reduction of the large Fermi surface to Fermi arc(area in momentum space that remains un-gapped) [17-19,29-41] as in Wyle/Dirac semi-metal. The exact nature of the pseudo-gap still remains an open issue.

In the past, Nayak et al.[ 33-35 ] have put forward $d$ density wave (DDW) order as the one for the pseudo gap (PG)phase of the hole-doped cuprates. This order [33-35] corresponds to spontaneous currents along the bonds of a square lattice for the ordering wave vector $\mathbf{Q}=(\pm\pi,\pm\pi)$. The DDW state preserves the combined effect of the microscopic time reversal symmetry violation and the translation by a lattice spacing. The net result is that the staggered magnetic flux produced by these currents is zero on the macroscopic scale. The formulation of Nayak et al.[33-35]involved d-wave superconductivity(DSC).The reason for the choice is that the d-wave superconducting symmetry and strong Coulomb repulsion are compatible :For strong on-site Coulomb repulsion the superconducting state needs to avoid same-site pairing which corresponds to isotropic k-space pairing. At the same time, the Coulomb repulsion generally favours anti-ferromagnetic tendencies and thus a spin-singlet superconducting state. The spin singlet state with the lowest number of nodes but still avoiding same-site pairing is exactly the d-wave state. The non-zero DSC gap $\Delta_k^{(sc)}$ requires an appropriate attractive interaction. The gap is, formally similar to the weak- coupling BCS gap equation, given by $\Delta_k^{(sc)}=\Delta_0(T)(\cos k_x a – \cos k_y a)$.The Chiral DDW (CDDW) order, which is a complex variant of the DDW order, was subsequently advocated for by S. Das Sarma et al. [36-39] as the one suitable for the PG phases ordering. The reason being it offers a theoretical explanation [36-39] of the non-zero polar Kerr effect observed in YBCO by Kapitulniket al.[39]. A very pertinent theoretical reason for the identification of the PG state with the CDDW state rather than with the DDW state is that the imaginary part of the d+id-wave order parameter $D_k=(-\chi_k+i\Delta_k)$ breaks the parity and the time reversal symmetry (TRS)of the normal state. The quantities $(\chi_k, \Delta_k)$ are given by [36-39] $\chi_k= -(\chi_0/2) \sin(k_x a) \sin(k_y a), \Delta_k= (\Delta_0^{(PG)}(T)/2)(\cos k_x a – \cos k_y a)$, and $\mathbf{k} = ( k_x, k_y )$. The normal state dispersion is imperfectly nested here . The corresponding evidence is that the onset of CDDW ordering leads to a peak in the anomalous Nernst signal

(ANS)[36,37]. The main contribution to this chirality induced ANS comes from the points ($\pm\pi(1-\varphi)$, $\pm\pi\varphi$),($\pm\pi\varphi$, $\pm\pi(1-\varphi)$) with $\varphi \sim 0.2258$ located roughly on the boundary of the Fermi pockets in the momentum space( cf. ref. [36,37]). Thus, in order to present a suitable description of cuprates in the framework of CDDW ordering, we have to turn our attention to this imperfectly nested dispersion. All energies are expressed in units of the first neighbour hopping. The second-neighbour hopping in the dispersion, which is known to be important for cuprates [33-44], frustrates the kinetic energy of electrons.

It is quite intriguing that, in most of the conventional superconductors(CSC) where BCS theory works, the pseudo-gap has not been observed as the predecessor state. In CSCs, the precursor state is generally a metal. In unconventional high temperature superconductor (UHTSC), however, initially undoped cuprates remain in an insulating, antiferromagnetic state known as a Mott insulator, with neighbouring electrons having opposite magnetic spins. Upon chemical doping, there is emergence of Fermi arc (area in momentum space that remains un-gapped) as in Wyle semi-metal. By lowering the temperature T from T*, at a given doping in the underdoped region the pseudo-gapped phase sets in. The flow without resistance is initiated for the (finite momentum) Cooper pairs formed close to the Fermi surface (the boundary between the highest occupied energy level and the next is known as the Fermi surface) at $T<T_c$.. The dominant interaction coupling the superconducting electron pairs, conjecturally, is none other than the same "super-exchange" spin coupling which causes the undoped cuprates to be antiferromagnetic Mott insulators. The reason for this surmise is that in this strongly correlated system, apart from interaction and orbital overlap, we have spin-orbit coupling(SOC) and its variant, namely spin-spin interaction. The SOC and its exotic effect will be included in sections 2 and 3. In this paper, since we do not consider superconductivity, no attractive interaction conjecture will be required. We note that degree of complexities in UHTSC is much higher than in a CSC.

The paper is organized as follows: In section 2 we derive an expression for the single-particle excitation spectrum in CDDW state to show the spin-momentum locking. In section 3, we investigate the spin texture and Dzyaloshinskii–Moriya interaction (DMI) coefficient in Bi2212 with our model Hamiltonian. The paper ends with brief discussion and conclusion in section 4.

**2. Pseudo-gap phase with interlayer tunnelling and Rashba coupling**

Motivated by the findings of S. Das Sarma et al.**[36-39]**, we assume that the CDDW order represents the pseudo-gap(PG) phase. We assume the momentum dependence of the pairing interactions required for this kind of ordering is given by the functions of the form $U_{x2-y2}(k,k') = U_1 (\cos k_x a – \cos k_y a)(\cos k'_x a – \cos k'_y a)$, $U_{xy}(k,k') = U_2 \sin(k_x a) \sin(k_y a) \sin(k'_x a) \sin(k'_y a)$, where $U_1$ and $U_2$ ($U_1 > U_2$) are the coupling strengths, and ($k_x, k_y$) belong to the first Brillouin zone (BZ). Furthermore, the cuprates (the special class of high-temperature superconductors (HTSCs)) consist of superconducting $CuO_2$ layers separated by spacer layers as in Bi2212. The simplest vertical hop is straight up via $t_b$. The bilayer split bands in Bi2212 display considerable dispersion with k: It corresponds to $[ -2t_b(c_x(a) - c_y(a))^2 ]$ where $t_b$ is an effective parameter for hopping within a single bilayer, i.e. it controls the intracell bilayer splitting and $c_j(a) \equiv \cos(k_j a)$ **[45]**. The compound Bi2212, however, involves intercell hopping ($t_z$), too. The total dispersion can now be written as

$\epsilon(\mathbf{k},k_z) = \varepsilon_\mathbf{k} - \mu + \varepsilon_{k_z}(\mathbf{k})$, where $\mathbf{k}$ and $k_z$ respectively denote the in-plane and out-of-plane components of $\mathbf{K} = (\mathbf{k},k_z)$. The dispersion $\varepsilon_\mathbf{k}$ has the usual form

$$\varepsilon_k = -2t\left(c_x(a) + c_y(a)\right) + 4t'c_x(a)c_y(a) - 2t''\left(c_x(2a) + c_y(2a)\right) - 4t'''(c_x(a)c_y(2a)$$

$$+ c_y(a)c_x(2a)) \qquad (1)$$

including the first, the second and the third neighbour hops due to the non-zero neighbour hopping and $\mu$ is the chemical potential of the fermion number. For simplicity, we have set the lattice constant equal to one. The term $\varepsilon_\mathbf{k}$ is the model dispersion associated with a single $CuO_2$ plane if the effects of $k_z$-dispersion are totally neglected. Early theories only took t into account, but the consistent results of local-density approximation, band-structure calculations [17-19] and angle-resolved photoemission spectroscopy (for over-doped, stripe-free materials) [17-19], have led to the usage of including also t', with t'/t = 0.1 for $La_2CuO_4$ and t'/t = 0.3 for $YBa_2Cu_3O_7$ and $Bi_2Sr_2Ca Cu_2O_8$, whereby the constant-energy contours of the expression for $\varepsilon_\mathbf{k}$ become rounded squares oriented in the [11]- and [10]-directions respectively. For the hole-doped materials, t' > 0 (for the electron-doped materials t' < 0), and, in all cases, t' < (t/2). The $\varepsilon_{k_z}(\mathbf{k})$ term accounts for the effect of coupling between different $CuO_2$ planes, and possesses a very different form in a single layer cuprate such as LSCO or NCCO than when a $CuO_2$ bilayer is present as in Bi2212. In the case of Bi2212[46],

$$\varepsilon_{k_z}(\mathbf{k}) = -\Gamma_z(\mathbf{k}, c_z(c/2))[(c_x(a) - c_y(a))^2/4 + a_0] \qquad (2)$$

where c denotes the lattice constant along the z-axis, and

$$\Gamma_z(\mathbf{k}, c_z(c/2)) = \pm (t_b^2 + A'^2 + 2t_bA'c_z(c/2))^{1/2}, A' = 4\ t_zc_x(a/2)c_y(a/2). \qquad (3)$$

The plus (minus) sign refers to the bonding (anti-bonding) solution and the term $t_z$ corresponds to the intercell hopping. The term $c_z$ arises because supposedly we have an infinite number of stacked layers. The term $A' = 4\ t_zc_x(a/2)c_y(a/2)$ is zero along the high symmetry line $X(\pi,0)$ —$M(\pi, \pi)$. This leads to a lack of $k_z$-dispersion along this high-symmetry line. The additional hopping $a_0$ allows for the presence of a splitting at $\Gamma(0,0)$. It is reported [47] that adequate control of the interlayer spacing albeit the interlayer hopping in Bi-based superconductors is possible through the intercalation of guest molecules between the layers. This could be a way to tune the hopping parameter. The coupling $t_b$ within a bilayer and the intercell coupling $t_z$ are both quite substantial as are the in-plane hopping terms beyond the NN term. It may be mentioned that Vishwanath et al.[1] had found that a spin-momentum locking that allows states of opposite spin to be localized in different parts of the unit cell. The spin-momentum locking, to be obtained in this section, which is similar to their observation in spirit save for the fact that the Wigner-Seitz cell is replaced by the brillouin zone. We shall see that when the Rashba-coupling is introduced together with the full form of the tunnelling matrix given by Eq.(2) and (3), while the nodal region of the momentum space will give rise to a spin-down hole current, the anti-nodal region will give rise to spin-up electron current in the bonding case. In the anti-bonding case, the result will be similar.

Suppose now $d^{(m)}_{k,\sigma}$ ( $\sigma = \pm 1$ for the real spin ) corresponds to the fermion annihilation operator for the single-particle state $(\mathbf{k},\sigma)$ in the layer m (m=1,2) of the system. As regards the interactions, $U \sum_i d^\dagger_{i\uparrow}d_{i\uparrow}d^\dagger_{i\downarrow}d_{i\downarrow}$ is the onsite repulsion of $d$ electrons, where the intra-layer $d$ electrons are locally interacting via a Hubbard-$U$ repulsion. We have not considered this term assuming the correlation effect are marginally relevant. In the basis $(d^{\dagger(1)}_{k,,\sigma}, d^{\dagger(1)}_{k+Q,\sigma}, d^{\dagger(2)}_{k,-\sigma}, d^{\dagger(2)}_{k+Q,-\sigma},)^T$, we consider the following bilayer Hamiltonian in momentum space :

$$H(k_x, k_y, k_z) = \begin{pmatrix} \varepsilon_k & D^\dagger_k & \tau(\mathbf{k}, k_z) & 0 \\ D_k & \varepsilon_{k+Q} & 0 & 0 \\ \tau^*(\mathbf{k}, k_z) & 0 & \varepsilon_k & D^\dagger_k \\ 0 & 0 & D_k & \varepsilon_{k+Q} \end{pmatrix} \quad (4)$$

where $\tau(\mathbf{k}, k_z) = \varepsilon_{k_z}(\mathbf{k}) + \alpha_k, \alpha_k = \alpha_0(-i\sin(k_x a) - \sin(k_y a))$ is the polarizing field(assumed to be in the z direction) led(Rashba) spin-orbit coupling. The quantity $\alpha_0$ is the coupling strength which is proportional to the strength of the field. Therefore, it is tunable. Here, as before, we assume the ordering wave vector $\mathbf{Q}=(Q_1=0.7742\pi, Q_2=0.2258\pi)$. We assume here the occurrence of spin-flip due to Rashba coupling $\alpha_k$. Together with the lifting of the spin-degeneracy, indication of spin-momentum locking in limited region of BZ is then expected.

We find from above that $H(k_x, k_y, k_z)$ is inversion asymmetric. The reason for the broken mirror reflection symmetry, apart from the presence of the Rashba coupling, is $D_k(k_x \to -k_x, k_y \to k_y) \neq D_k(k_x, k_y)$ due to the presence of chirality(d+id). For the non-chiral system(id), however, $D_k(k_x \to -k_x, k_y \to k_y) = D_k(k_x, k_y)$, i.e. the inversion symmetry is preserved. Note that the Hamiltonian in (4) violates time reversal symmetry(TRS) also. Never-the-less, as we see below, we get real eigenvalues in BZ. Upon ignoring $\tau(\mathbf{k}, k_z)$ altogether, the energy eigenvalues in the CDDW case are given by multiple roots. These are $E^{(\alpha)}(\mathbf{k}) = \varepsilon_k^U + \alpha\Delta_k$ where $\alpha = \pm 1$, $\varepsilon_k^U = (\varepsilon_k + \varepsilon_{k+Q})/2, \varepsilon_k^L = (\varepsilon_k - \varepsilon_{k+Q})/2$, and $\Delta_k = \left(\varepsilon_k^{L^2} + |D_k|^2\right)^{\frac{1}{2}}$. The plots of free electron dispersion and, once the CDDW order sets in. are shown in Figure1. We find that the valence and the conduction bands corresponding to free electrons are partially full(see 1(a) and 1(b)). Therefore, the system conducts. Once the CDDW order sets in, the pseudo-gap phase displays a nodal-anti-nodal dichotomous feature, i.e. excitations with infinite lifetimes(say, in a Hartee-Fock treatment of the e-e repulsion), have un-gapped nodal points and maximally gapped anti-nodal points (see 1(c) and (d)). The reason basically lies in the particle-hole asymmetry of the excitation spectrum $\varepsilon_1, \varepsilon_2 = \varepsilon_k^U \pm \sqrt{\varepsilon_k^{L^2} + D^\dagger_k D_k}, \varepsilon_k^U = \frac{\varepsilon_k + \varepsilon_{k+Q}}{2}$, and $\varepsilon_k^L = \frac{\varepsilon_k - \varepsilon_{k+Q}}{2}$. With $Q_1 = 0.7742$pi, $Q_2 = 0.2258$pi,$\varepsilon_k^U \sim 0$, and $\varepsilon_k^L$ is minimum (maximum) at the nodal(antinodal)point.

In the case of the perfectly nested dispersion ($\varepsilon_k = -\varepsilon_{k+Q}$), it is easy to see that the eigenvalue equation which is a quartic may be written as $\varepsilon^4 - 2 \varepsilon^2 b - 4 \varepsilon c + d = 0$ or,

$$\varepsilon^4 - 2\varepsilon^2 b + b^2 = 4\varepsilon c + b^2 - d, \qquad (5)$$

where $b = \frac{1}{2}(t_k^2 + 2(\varepsilon_k^2 + |D_k|^2))$, $c = \frac{1}{2}\varepsilon_k t_k^2$, and $d = (\varepsilon_k^2 + |D_k|^2)^2 - t_k^2 \varepsilon_k^2$. For simplicity, we have replaced here $\tau(\mathbf{k}, k_z)$ by $t_k$. We now add and subtract an as yet unknown variable z within the squared term $(\varepsilon^4 - 2\varepsilon^2 b + b^2)$: $(\varepsilon^2 - b + z - z)^2 = 4\varepsilon c + b^2 - d$, or $(\varepsilon^2 - b + z)^2 = 2z\varepsilon^2 + 4\varepsilon c + (z^2 - 2bz + b^2 - d)$. Upon ignoring the term $(4\varepsilon c)$ we shall get a bi-quadratic with values of 'ε' given by $\varepsilon^2 \approx b \pm \sqrt{(b^2 - d)}$. We shall see below that without the term $(4\varepsilon c)$ a discussion of the spin-texture, etc., which is one of our tasks here, does not seem to be possible. The left-hand side of Eq.(5) is a perfect square in ε. Therefore, we need to rewrite the right hand side in that form as well. For this we require that the discriminant of the quadratic in the variable ε to be zero. This yields $16c^2 - 8z(z^2 - 2bz + b^2 - d) = 0$ or, $z^3 - 2bz^2 + (b^2 - d)z - 2c^2 = 0$. The corresponding depressed cubic equation $s^3 + ps + q = 0$ has the discriminant function $D = -4p^3 - 27q^2$, where $s = z - (2b/3)$, $p = -(b^2/3 + d)$, and $q = 2(b^3/27 - bd/3 - c^2)$. Since we find D negative in the entire Brillouin zone, the equation $s^3 + ps + q = 0$ has one real root and two complex conjugate roots. Suppose we denote this root by $s_0(a\mathbf{k})$, then the corresponding 'z' will be denoted by $z_0(a\mathbf{k}) = s_0(a\mathbf{k}) + 2b/3$. After lengthy algebra we obtain real $z_0 = 2b/3 + (-q/2 + 0.5\psi^{1/2})^{1/3} + (-q/2 - 0.5\psi^{1/2})^{1/3}$, where $\psi = q^2 + 4(b^3/9 + d/3)^3$. Using (5) one may then write $\varepsilon^2 = b - z_0 \pm \{\sqrt{(2z_0)}\varepsilon + \sqrt{(2/z_0)}c\}$ or, $\varepsilon^2 \mp \sqrt{(2z_0)}\varepsilon + (-b + z_0 \mp c\sqrt{(2/z_0)}) = 0$. These equations are expected to yield the band structure. We have not been able to find a reasonably sized window in momentum space where eigenvalues would be real. The obvious conclusion is that the assumption of the nested dispersion and the replacement $\tau(\mathbf{k}, k_z)$ by $t_k$ do not work. In what follows we, therefore, proceed in a straightforward manner without any assumptions.

The eigen values of the matrix in (4) are, once again, given by the quartic

$$A\varepsilon^4 + B(k)\varepsilon^3 + C(k)\varepsilon^2 + D(k)\varepsilon + E = 0,$$

where

$$A = 1, \quad B(k) = -4\varepsilon_k^U, \quad C(k) = 4\varepsilon_k^{U^2} + 2\gamma_1^2 - \gamma_2^2 - \gamma_3^2,$$

$$D(k) = -4\varepsilon_k^U \gamma_1^2 + 2\varepsilon_{k+Q}\gamma_2^2 + 2\varepsilon_k \gamma_3^2, \quad E(k) = \gamma_1^4 - \gamma_2^2 \varepsilon_{k+Q}^2 - \gamma_3^2 \varepsilon_k^2 + \gamma_0^4,$$

$$\varepsilon_k^U = \frac{\varepsilon_k + \varepsilon_{k+Q}}{2}, \quad \varepsilon_k^L = \frac{\varepsilon_k - \varepsilon_{k+Q}}{2}, \quad \gamma_1^2 = (\varepsilon_k \varepsilon_{k+Q} - |D_k|^2),$$

$$\gamma_2^2 = \{(\varepsilon_{k_z}(\mathbf{k}) - \alpha_0 \sin(k_y a))^2 + \alpha_0^2 \sin(k_x a)\},$$

$$\gamma_3^2 = \{(\varepsilon_{k_z}(\mathbf{k}+\mathbf{Q}) - \alpha_0 \sin(k_y a + Q_2))^2 + \alpha_0^2 \sin(k_x a + Q_1)\},$$

$$\gamma_0^4 = -2|D_k|^2 (\varepsilon_{k_z}(\mathbf{k}) - \alpha_0 \sin(k_y a))(\varepsilon_{k_z}(\mathbf{k}+\mathbf{Q}) - \alpha_0 \sin(k_y a + Q_2))$$

$$-2|D_k|^2 \alpha_0^2 \sin(k_x a + Q_1) \times \sin(k_x a) + \gamma_2^2 \gamma_3^2. \qquad (7)$$

In view of the Ferrari's solution of a quartic equation, we find the roots as

$$\epsilon_j(s,\sigma,k) = \sigma\sqrt{\frac{z_0(k)}{2}} + \varepsilon_k^U + s\left(b_0(k) - \left(\frac{z_0(k)}{2}\right) + \sigma c_0(k)\sqrt{\frac{2}{z_0(k)}}\right)^{\frac{1}{2}}, \quad (8)$$

where j= 1,2,3,4, $\sigma = \pm 1$ is the spin index and s = $\pm 1$ is the band-index. The other functions appearing in (3) are defined below:

$$z_0(k) = \frac{2b_0(k)}{3} + \left(\frac{1}{2}\Delta^{\frac{1}{2}}(k) - A_0(k)\right)^{\frac{1}{3}} - \left(\frac{1}{2}\Delta^{\frac{1}{2}}(k) + A_0(k)\right)^{\frac{1}{3}}, (9)$$

$$A_0(k) = \left(\frac{b_0^3(k)}{27} - \frac{b_0(k)d_0(k)}{3} - c_0^2(k)\right), b_0(k) = \frac{3B^2(k) - 8C(k)}{16}, c_0(k) = \frac{-B^3(k) + 4B(k)C(k) - 8D(k)}{32}, \quad (10)$$

$$d_0(k) = \frac{-3B^4(k) + 256E(k) - 64B(k)D(k) + 16B^2(k)C(k)}{256}, \quad (11)$$

$$\Delta(k) = \left(\frac{8}{729}b_0^6 + \frac{16d_0^2 b_0^2}{27} + 4c_0^4 - \frac{4d_0 b_0^4}{81} - \frac{8c_0^2 b_0^3}{27} + \frac{8c_0^2 b_0 d_0}{3} + \frac{4}{27}d_0^3\right), \quad (12)$$

The plots of bands in (8) in the bonding/ anti-bonding cases belonging to **the nodal and anti-nodal regions** are shown in Figures 2. The numerical values of the parameters to be used in the calculation are t =1 , t′/t = −0.28(hole-doping), t ″/t = 0.1, t ‴/t = 0.06, $t_b/t$ = 0.3, $t_z/t$ = 0.1, $\frac{a_0}{t}$ = 0.53, and $a_0$ = 0.4. Throughout the whole paper, we choose *t* to be the unit of energy. In Figure 2(a) we have a plot of quasi-particle excitation (QP) spectrum given by Eq.(8) in the bonding case as function of dimensionless momentum $k_x a$ for $k_y a = \pi/2$ and $k_z a = \pi$. Since spin-down valence and conduction bands are partially empty , the spin-down QP conduction is possible. A plot of QPsin the bonding case, as a function of dimensionless momentum $k_x a$ for $k_y a = \pi$ and $k_z a = \pi$ is given in Figure (b). The plot displays band crossing and huge spectral gap at the high symmetry point R($\pi,\pi,\pi$).A plot of quasi-particle excitation spectrum in the anti-bonding case as function of $k_x a$ for $k_y a = 0$ and $k_z a = 0$ is shown in Figure (c). Since the spin-up conduction band is partially empty, the spin-up electron conduction is possible.A band pair of opposite spins, where one of the partners is a partially empty band and the other is full band, is almost coinciding in energy with zero density of states(DOS) at their meeting point in momentum space. In Figure 2(d) we have a plot of QP excitation spectrum in the anti-bonding case as function of dimensionless momentum $k_x a$ for $k_y a = \pi/2$ and $k_z a = 0$. Once again, since the spin-down valence band is partially empty , the spin-down hole conduction is possible.Thus, the spin-momentum locking is evident, While the nodal region of the momentum space can give rise to a spin-down hole current, the anti-nodal region can give rise to spin-up electron current. We find that the common ground between two types of SMLs, corresponding to the bonding and the anti-bonding cases, is that the states of opposite spin are to be found in different parts of the Brillouin zone. On the experimental front, recently Vishwanath et al.[1] discovered that Bi2212, has a nontrivial spin texture with the spin-momentum locking. They used spin- and angle-resolved photoemission spectroscopic technique to unravel this fact. We shall discuss briefly below this feature.

One can gap out the helical edge states by introducing a Zeeman term that explicitly breaks the protecting time-reversal symmetry. As we have seen above in Eq. (8), we obtain a term $\sqrt{\frac{z_0(k)}{2}}$ which has different sign for opposite spins and connection with momentum (see Figure-3).Usually, the effects of an external magnetic field, B, perpendicular to the CuO$_2$plane may be captured by two additional terms in the planar Hamiltonian. The first term describes the orbital coupling to magnetic field through the minimal coupling $k_x \to k_x + (e/c)A_x$, where one may choose **A** = [−By, 0] is the vector potential in the Landau gauge. The second term describes the coupling of the spins to the magnetic field and is given by the Zeeman contribution. Since, the spin index $\sigma$ occurs twice in Eq. (8)(See $\epsilon_j(s,\sigma,k) = \sigma\sqrt{\frac{z_0(k)}{2}} + \varepsilon_k^U + s\left(b_0(k) - \left(\frac{z_0(k)}{2}\right) + \sigma c_0(k)\sqrt{\frac{2}{z_0(k)}}\right)^{\frac{1}{2}}$, where the index is highlighted by red ink.), the term $\sqrt{\frac{z_0(k)}{2}}$ in question does not act like magnetic energy.

## 3. Spin Texture

The spin texture of the surface states in topological insulator(TI) forms due to SOC. For the Bi2212 system also it is expected that Rashba SOC will induce this texture. Therefore, we now focus on this aspect. The **spin texture** s(n,**k**) is defined as the expectation value of a vector operator **S**$_j$= I$_{2\times 2}\otimes \boldsymbol{\sigma}_j$where $\boldsymbol{\sigma}_j$ are Pauli matrices on a two dimensional **k**-grid and $\otimes$ stands for the tensor product. At **k** for the state n (or nth band) it is defined as an expectation value **s** z(n,**k**) = $\langle S_z \rangle^{(n)}=\langle n|S_z|n\rangle$. Obviously enough, to calculate this we need eigenvectors of the Hamiltonian matrix (4) for an eigenvalue $E_n$. This complete set is given by

$$|\Psi_n\rangle = \varsigma_n^{-1}(\mathbf{k},k_z)\begin{pmatrix} 1 \\ \frac{D_k}{E_n-\varepsilon_{k+Q}} \\ \frac{\tau^*(\mathbf{k},k_z)(E_n-\varepsilon_{k+Q})}{[(E_n-\varepsilon_{k+Q})(E_n-\varepsilon_k)-|D_k|^2]} \\ \frac{\tau^*(\mathbf{k},k_z)D_k}{[(E_n-\varepsilon_{k+Q})(E_n-\varepsilon_k)-|D_k|^2]} \end{pmatrix}, \quad n=1,2,3,4, \quad (13)$$

$$\varsigma_n(\mathbf{k},k_z) = \sqrt{\left(\frac{(|D_k|^2)}{(E_n-\varepsilon_{k+Q})^2} + \frac{(|\tau(\mathbf{k},k_z)|^2)(E_n-\varepsilon_{k+Q})^2}{(E_n-\varepsilon_1)^2(E_n-\varepsilon_2)^2} + \frac{(|\tau(\mathbf{k},k_z)|^2)(|D_k|^2)}{(E_n-\varepsilon_1)^2(E_n-\varepsilon_2)^2}\right)} \quad (14)$$

$$\tau(\mathbf{k},k_z) = \varepsilon_{k_z}(\mathbf{k}) + \alpha_k, \alpha_k = \alpha_0(i\sin(k_xa) - \sin(k_ya)) \quad (15)$$

where the full expression for $\varepsilon_{k_z}(\mathbf{k})$ could be found in Eqs.(2) and (3). Upon using (13)-(15), the spin textures are obtained in a straightforward manner. The contour plots are shown in Figure 3 for various values of $a_0$ and $\alpha_0$ with $ak_z=\pi$. The texture is found over a large range of parameters in this material with Rashba SOC and broken surface inversion starting with a critical value of $\alpha_0$.

We have found this critical value to be 0,35 for $a_0 = 0.40$. For $\alpha_0 < 0.35$, the expectation value turns out to be complex. Most importantly, as can be seen from Figure 3 that, the texture or the expectation value spreads over the larger region of the Brillouin zone as $\alpha_0$ (and $a_0$) increases. This means the texture is tunable by electric field(and by intercalation[47]). Upon using (13)-(15) we also obtain the values of ($\langle n|S_x n\rangle - \langle n|S_y|n\rangle$)spin texture in the bonding case in the $ak_x$–$ak_y$ plane for $ak_z = \pi$(see Figure 4) with the polarising field $\alpha_0$ in the z direction. The plots are for different values of $a_0$ and $\alpha_0$. The band index 'n' stands for the spin down bands(see Figure 2(a)) $\in (\downarrow, \sigma = \pm 1, k) = -\sqrt{\frac{z_0(k)}{2}} + \varepsilon_k^U \pm \left(b_0(k) - \left(\frac{z_0(k)}{2}\right) - c_0(k)\sqrt{\frac{2}{z_0(k)}}\right)^{\frac{1}{2}}$ intersecting with Fermi energy. The entire figure plane is rough-hewn and bumpy with quasi-periodic, tapering expectation value $\langle S_z\rangle^{(n)}$ or ($\langle n|S_x n\rangle - \langle n|S_y|n\rangle$) spin textures.

The Dzyaloshinskii–Moriya interaction (DMI) [48] is considered as one of the most important energies for specific chiral textures such as magnetic skyrmions. The keys of generating DMI are the absence of structural inversion symmetry and energy corresponding to spin–orbit coupling. What is commonly understood by DMI is an antisymmetric exchange interaction that can be written for two spin moments as $\mathbf{D}_{12} \cdot (\mathbf{S}_1 \times \mathbf{S}_2)$, where $\mathbf{D}_{12}$ is referred to as the interaction vector. In a bid to initiate the investigation of this problem we consider the operator $\hat{O}_{\mu\nu}(k) = \widehat{v_\mu}\Sigma_\nu + \Sigma_\nu \widehat{v_\mu}$ where $\Sigma_j = \begin{pmatrix} \sigma_j & 0 \\ 0 & \sigma_j \end{pmatrix}$, and $\sigma_j$ are the Pauli matrices. Here the velocity operator in momentum eigenbasis is given by $\hat{v}_j = \left(\frac{1}{\hbar}\right)\frac{\partial \hat{H}}{\partial k_j}$; $\hat{H}$ being the full Hamiltonian given by the Hamiltonian density in Eq.(4). The properties of the Pauli and the four by four $\Sigma$ matrices are similar. The DMI coefficient $D_{\mu\nu}$ [48], which also corresponds to spin current $J_{\mu\nu}$, is defined as $D_{\mu\nu} = \sum_{k,\sigma}(\frac{1}{4})\langle \Psi^\dagger_{k,\sigma} \hat{O}_{\mu\nu}(k)(k)\Psi_{k,\sigma}\rangle$ is a thermally averaged expectation value. The row vector

$$\Psi^\dagger_{k,\sigma} = \left(d^{(1)\dagger}_{k,\sigma} d^{(1)\dagger}_{k+Q,\sigma} d^{(2)\dagger}_{k,-\sigma} d^{(2)\dagger}_{k+Q,-\sigma}\right).$$

In view of Eq.(4), the spin current density components $J_{\mu x}$ could be written as $J_{\mu x} =$

$$\sum_{k,\sigma}(\frac{1}{4\hbar})\langle \Psi^\dagger_{k,\sigma} \hat{O}_{\mu x}(k) \Psi_{k,\sigma}\rangle$$

where the operator $\hat{O}_{\mu x}(k)=$

$$\begin{pmatrix} \partial_{k_\mu}(D_k + D^\dagger_k) & \partial_{k_\mu}(\varepsilon_k + \varepsilon_{k+Q}) & \partial_{k_\mu}\tau(\mathbf{k}, k_z) & 0 \\ \partial_{k_\mu}(\varepsilon_k + \varepsilon_{k+Q}) & \partial_{k_\mu}(D_k + D^\dagger_k) & 0 & \partial_{k_\mu}\tau(\mathbf{k}, k_z) \\ \partial_{k_\mu}\tau^*(\mathbf{k}, k_z) & 0 & \partial_{k_\mu}(D_k + D^\dagger_k) & \partial_{k_\mu}(\varepsilon_k + \varepsilon_{k+Q}) \\ 0 & \partial_{k_\mu}\tau^*(\mathbf{k}, k_z) & \partial_{k_\mu}(\varepsilon_k + \varepsilon_{k+Q}) & \partial_{k_\mu}(D_k + D^\dagger_k) \end{pmatrix} \quad (16)$$

For the components ($J_{\mu x}$) in the momentum space, $\frac{\partial}{\partial k_\mu}(D^\dagger{}_k + D_k)$ is zero at the nodal and anti-nodal points; the terms $\frac{\partial}{\partial k_\mu}(\varepsilon_k + \varepsilon_{k+Q})$ and $\frac{\partial}{\partial k_\mu}\tau(\mathbf{k}, k_z)$ (appearing due to the presence of the z-axis dispersion and the Rashba spin-orbit coupling) are, however, non-zero. The thermal averages $\langle d^{\dagger(m)}_{k,\sigma} d^{(m)}_{k+Q,\sigma}\rangle, \langle d^{\dagger(m)}_{k,\sigma} d^{(n)}_{k+Q,-\sigma}\rangle$, etc. (m,n = 1,2,3,---------) need to be determined to obtain these components in full form. For this purpose, we shall proceed with the finite-temperature formalism here. Since the Hamiltonian is completely diagonal one can write down easily the equations for the operators $\{d^{(m)}_{k+Q,\sigma}(\tau), d^{\dagger(m)}_{k,\sigma}(\tau)\}$, where the imaginary time evolution an operator $O$ is given by $O(\tau)=\exp(H\tau)\,O\,\exp(-H\tau)$, to ensure that the thermal averages in the equations above are determined in a self-consistent manner. The Green's functions $G^{(m,m)}(k\sigma, k\sigma, \tau) = -\langle T_\tau\{d^{(m)}_{k,\sigma}(\tau)\,d^{\dagger(m)}_{k,\sigma}(0)\}\rangle$, etc., where $T_\tau$ is the time-ordering operator which arranges other operators from right to left in the ascending order of imaginary time $\tau$, are of primary interest. We find equation of motion of these functions. Upon taking Fourier transform (FT), the equations reduce to a set of simultaneous algebraic equations. Solving them we obtain the FTs which yield the direct expression for these functions when one takes inverse Fourier transform. The other components in momentum space, for example $J_{\mu y}$, will have the following expression:

$$\sum_{k,\sigma}\left(\frac{1}{4i\hbar}\right)\langle \Psi^\dagger_{k,\sigma}\,\hat{O}_{\mu y}(k)\,\Psi_{k,\sigma}\rangle$$

where the operator $\hat{O}_{\mu y}(k)=$

$$\begin{pmatrix} \partial_{k_\mu}(D_k - D^\dagger{}_k) & \partial_{k_\mu}(\varepsilon_k + \varepsilon_{k+Q}) & 0 & \partial_{k_\mu}\tau(\mathbf{k}, k_z) \\ -\partial_{k_\mu}(\varepsilon_k + \varepsilon_{k+Q}) & \partial_{k_\mu}(D_k - D^\dagger{}_k) & -\partial_{k_\mu}\tau(\mathbf{k}, k_z) & 0 \\ 0 & \partial_{k_\mu}\tau^*(\mathbf{k}, k_z) & \partial_{k_\mu}(D_k - D^\dagger{}_k) & \partial_{k_\mu}(\varepsilon_k + \varepsilon_{k+Q}) \\ -\partial_{k_\mu}\tau^*(\mathbf{k}, k_z) & 0 & -\partial_{k_\mu}(\varepsilon_k + \varepsilon_{k+Q}) & \partial_{k_\mu}(D_k - D^\dagger{}_k) \end{pmatrix} \quad (17)$$

One may write down $J_{\mu z}$ in the momentum space in the similar manner. We notice a striking result that indicates a special status of the anti-nodal point in the Brillouin zone (BZ): Seemingly, this is the only point in the BZ where the spin current density $J_{\mu y}= (J_{xy}, J_{yy}, J_{zy})$ will be a skew matrix. The physical significance of this result in the skyrmionic context (skyrmions exist in real space) is not yet clear to us, though the real space passage is achieved very easily using a complete set of basis set $|\psi_n\rangle$, where the band index is 'n'. In view of the completeness condition, one can write for the arbitrary state vectors $|\Phi\rangle = \hat{O}|\varphi\rangle = \hat{O}\sum_n |\psi_n\rangle\langle\psi_n|\varphi\rangle$. The operator $\hat{O}_{\mu\nu}(k) = \widehat{v_\mu}\Sigma_\nu + \Sigma_\nu \widehat{v_\mu}$ as above. We assume that the inner product $\langle\psi_n|\varphi\rangle$ is dependent on band index 'n' and is given by

$$\phi_n(x) = (2/d)^{1/2}\left[\sin\left\{\frac{n\pi\left(x+\frac{d}{2}\right)}{d}\right\}\right]. \quad (18)$$

($n = 1, 2, 3, 4$). Since there are several examples of complete orthogonal systems, viz., the Legendre polynomials over $[-1,1]$, and Bessel function over $[0,1]$, and so on, one may ask why show preference for $\{\sin(nx),\cos(nx)\}$ over $[-\pi, \pi]$. The reason is—this is simpler to handle and it replicates the model of a periodic crystal potential, terminating at the surface, where it undergoes a jump abruptly to the vacuum level. The assumption is also guided by the fact that a plane of the Bi2212 bilayer system is the x-y plane with $-\infty \leq y \leq \infty$ and $-\frac{d}{2} \leq x \leq \frac{d}{2}$. This ensures $\phi_n\left(x = -\frac{d}{2}, +\frac{d}{2}\right) = 0$. The expectation value $\langle\varphi|\widehat{O}|\varphi\rangle$ with arbitrary state vector is equated with the thermal expectation value. We obtain

$$\langle\varphi|\widehat{O}|\varphi\rangle = \sum_n \langle\varphi|\hat{O}|\psi_n\rangle \phi_n(x) = \sum_{m,n} \phi_m(x)\langle\psi_m|\hat{O}|\psi_n\rangle \phi_n(x) =$$
$$\left(\frac{2}{d}\right)\sum_{m,n} \sin\left\{\frac{m\pi\left(x+\frac{d}{2}\right)}{d}\right\} \langle\varphi_m|\hat{O}|\varphi_n\rangle \sin\left\{\frac{n\pi\left(x+\frac{d}{2}\right)}{d}\right\}. \quad (19)$$

Here the expectation value $\langle\varphi_m|\hat{O}|\varphi_n\rangle$ is $\langle\Psi^\dagger_{k,\sigma}\hat{O}_{\mu y}(k_\mu \to -i\partial_\mu)\Psi_{k,\sigma}\rangle$ given by (17). In view of the preceeding discussion, the in-plane component of the spin-current density ($J_x$) in real space associated with the velocity operator $\widehat{v_x}$ is given by

$$J_{xy}(x, k_y, k_z, \sigma) = \left(\frac{2}{d}\right)\sum_{m,n} \sin\left\{\frac{m\pi\left(x+\frac{d}{2}\right)}{d}\right\} \langle\Psi^\dagger_{k,\sigma}\hat{O}_{xy}(k_x \to -i\partial_x)\Psi_{k,\sigma}\rangle \sin\left\{\frac{n\pi\left(x+\frac{d}{2}\right)}{d}\right\},$$

plus the two other terms

$$J_{xx}(x, k_y, k_z, \sigma) = \left(\frac{2}{d}\right)\sum_{m,n} \sin\left\{\frac{m\pi\left(x+\frac{d}{2}\right)}{d}\right\} \langle\Psi^\dagger_{k,\sigma}\hat{O}_{xx}(k_x \to -i\partial_x)\Psi_{k,\sigma}\rangle \sin\left\{\frac{n\pi\left(x+\frac{d}{2}\right)}{d}\right\},$$

$$J_{xz}(x, k_y, k_z, \sigma) = \left(\frac{2}{d}\right)\sum_{m,n} \sin\left\{\frac{m\pi\left(x+\frac{d}{2}\right)}{d}\right\} \langle\Psi^\dagger_{k,\sigma}\hat{O}_{xz}(k_x \to -i\partial_x)\Psi_{k,\sigma}\rangle \sin\left\{\frac{n\pi\left(x+\frac{d}{2}\right)}{d}\right\}. \quad (20)$$

The total spin-current density component could be obtained by integration:

$$\left(\frac{2}{d}\right)\sum_{m,n,\mu} \int_{-d/2}^{d/2} dx \sin\left\{\frac{m\pi\left(x+\frac{d}{2}\right)}{d}\right\} \langle\Psi^\dagger_{k,\sigma}\hat{O}_{x\mu}(k_x \to -i\partial_x)\Psi_{k,\sigma}\rangle \sin\left\{\frac{n\pi\left(x+\frac{d}{2}\right)}{d}\right\}. \quad (21)$$

The spin-current itself, in real space, associated with the velocity operator $\widehat{v_x}$ is given by

$$\sum_{k,\sigma}\left(\frac{1}{2i\hbar d}\right)\sum_{m,n,\mu} \int_{-d/2}^{d/2} dx \sin\left\{\frac{m\pi\left(x+\frac{d}{2}\right)}{d}\right\} \langle\Psi^\dagger_{k,\sigma}\hat{O}_{x\mu}(k_x \to -i\partial_x)\Psi_{k,\sigma}\rangle \sin\left\{\frac{n\pi\left(x+\frac{d}{2}\right)}{d}\right\}. \quad (22)$$

As noted in section 1, these outlines are expected to be useful in a full-throttle investigation of the Moriya vector **D**$_{ij}$ in similar lines as that of Moriya[1c] starting with the aforementioned Bloch Hamiltonian for Bi2212 bilayer system.

## 4. Discussion and Conclusion

Magnetic skyrmions are particle-like solitonic, complicated spin textures of topological origin, which exists in the real space. The intensive research activity on skyrmions has been driven bytheir potential applications to next-generation memory, logic, and neuromorphiccomputing devices. Thenon-centrosymmetric and centrosymmetric, both the categories of magnetic materials can host skyrmions. So far, the majority of experimentally identified skyrmion hosting materials are non-centrosymmetric except for a few centrosymmetric Gd-based compounds. The initiation of our investigation of the Moriya vector **D**$_{ij}$ is guided by the above-mentioned potential applications. There is yet another class of quasi-particle, viz. majoranas, with great potential in the field of computing. We note few relevant lines related to this problem regarding this particle below.

As in the Dirac system, to realize non-abelian anyons we may introduce the z-direction oriented TRS breaking Zeeman field ($H_z=\sigma g\mu_B B$) in the present coexisting PG + DSC(d wave superconductor with gap $\Delta^{sc}(k)=\Delta^{sc}_0 \sin(k_x a)\sin(k_y a)$). For the description of the PG+DSC state of this superconductor we may start with Gorkov type Hamiltonian in the basis $(d^\dagger_{k,\sigma} d^\dagger_{k,-\sigma} d_{-k,\sigma} d^\dagger_{k+Q,\sigma})^T$

$$\begin{pmatrix} E_k & \alpha^*_k & 0 & D^\dagger_k \\ \alpha_k & E_k & \Delta^{sc}(k) & 0 \\ 0 & \Delta^{sc}(k) & -E_{k+Q} & 0 \\ D_k & 0 & 0 & E_{k+Q} \end{pmatrix} \quad (23)$$

where $E_k = \varepsilon_k - \mu + \sigma g\mu_B B$. We find that (in the nodal region), as in a s-wave superconductor, for $B \geq B_c (B_c = \sqrt{\{(\mu-4t'')^2 + \Delta_0^{sc2} + \alpha_0^2 - \left(\frac{\chi_0}{2}\right)^2\}}/g\mu_B$, a vortex in this system becomes a non-Abelian anyon with a Majorana zero mode required for the fault tolerant quantum computation. The pseudo-gap being anathema to superconductivity is evident from the fact that $\Delta_0^{sc2}$ and $\left(\frac{\chi_0}{2}\right)^2$ occur with opposite signs in theexpression for $B_c$. We hope to present details in a separate communication.

In conclusion, the presence of Rashba spin–orbit coupling plays a key role on various aspects of spin transport. It can be detected in a material via magnetic field-induced quantum oscillations, electric-dipole spin resonance, and weak antilocalization. In the context of the present problem, we have witnessed the role it has played. The complete investigation of the DMI vector in real space is expected to unravel more of it. There are other problems, such as the Zitterbewegung effect, quantum anomalous and magnetoelectric effects, and Floquet physics, which need our attention. We hope to terminate this trail in a few years time with the investigations of the spin–orbit torque magnetic memory device, and the spin–orbit Qubit.

# Figures and Figure Captions

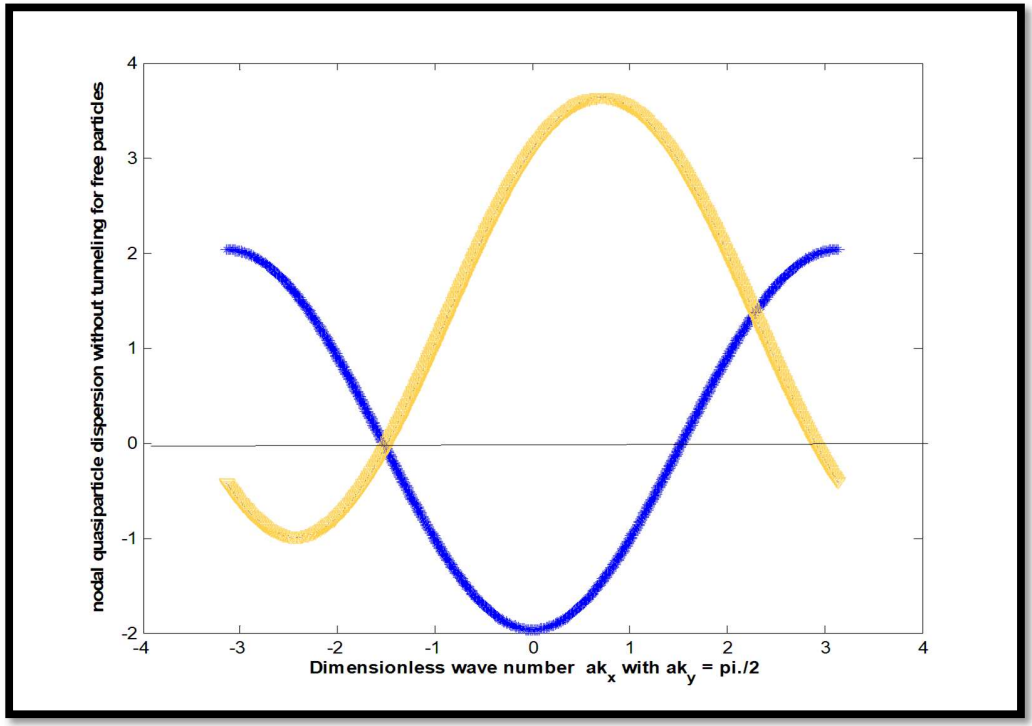

(a)

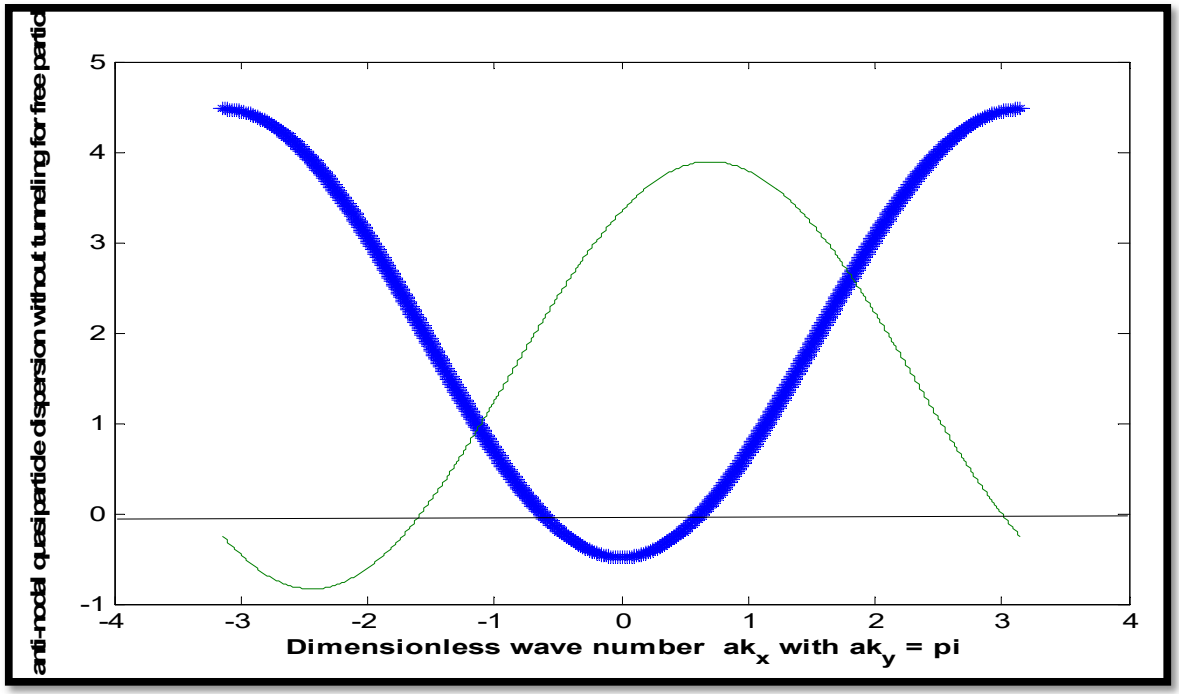

(b)

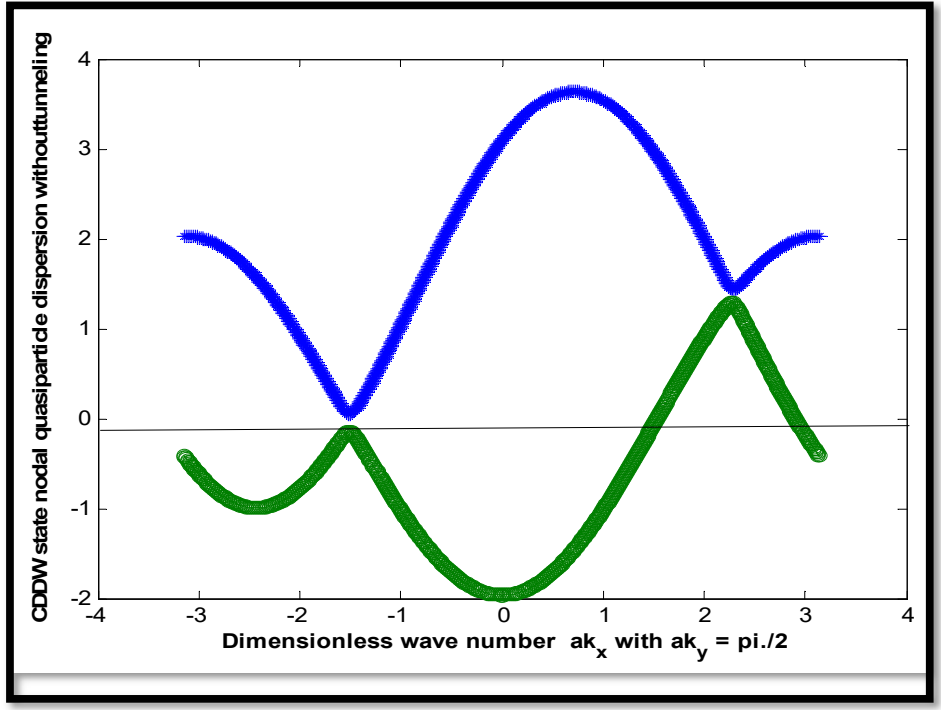

(c)

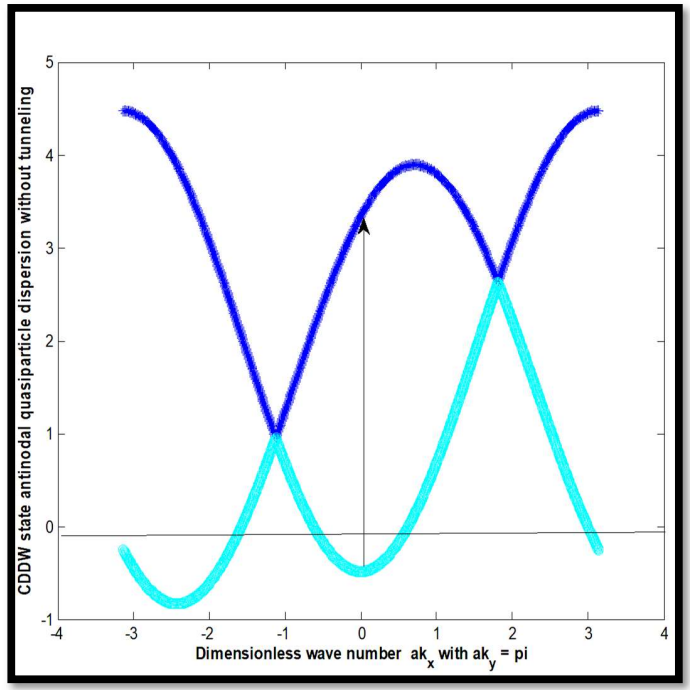

(d)

**Figure 1.** The two-band energy spectra of free electron dispersion and CDDW state ($\varepsilon_1, \varepsilon_2 = \varepsilon_k^U \pm \Delta_k$) without and with the interlayer tunneling at the nodal($ak_x \sim \frac{\pi}{2}, ak_y = \frac{\pi}{2}$) and the anti-nodal($ak_x \sim 0, ak_y = \pi$) regions. The chemical potential μ

represented by solid, horizontal line μ ~0 is located as shown. In figures (a) and (b), the free particle bands are partially empty and hence conduction is possible. From figure (c), which correspond to CDDW state nodal point excitations with or without tunneling, it may be seen that the some points in the momentum space are ungapped and therefore Fermi arcs are possible. However, in (d) there is a wide gap at the anti-node. The holes are conductive for (c) and (d). The parameter values are, μ = -0.035, $Q_1$ = 0.7742.pi, $Q_2$ = 0.2258.pi, $t = 1$, $t' = -0.12$, $t'' = 0.01$, , $t_0 = 0.005$, and $\Delta_0^{PG} = 0.01$.

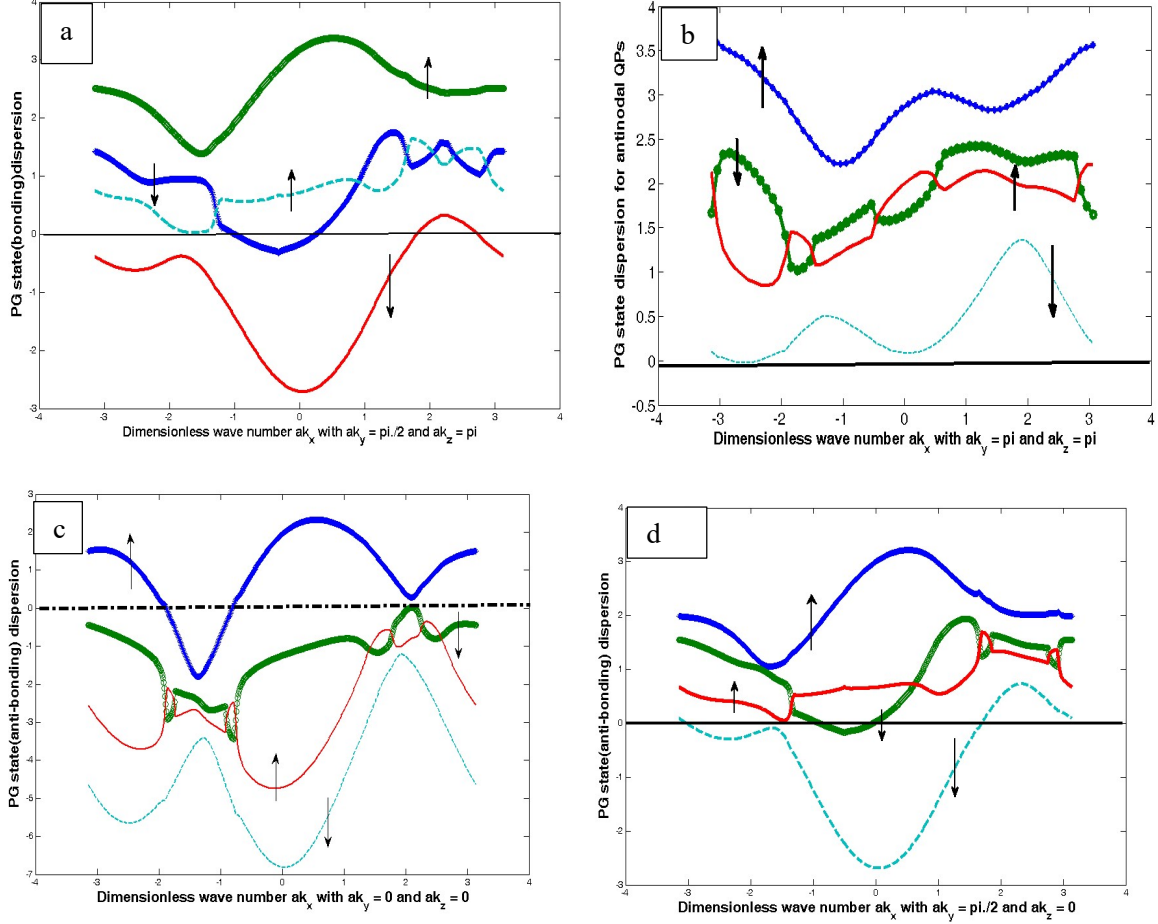

**Figure 2. (a)** A plot of quasi-particle excitation (QP) spectrum given by Eq.(8) in the bonding case as function of dimensionless momentum $k_x a$ for $k_y a = \pi/2$ and $k_z a = \pi$. The horizontal line represents the Fermi energy. Since spin-down valence and conduction bands are partially empty , the spin-down QP conduction is possible. **(b)** A plot of QPs in the bonding case, as a function of dimensionless momentum $k_x a$ for $k_y a = \pi$ and $k_z a = \pi$ .**(c)** A plot of quasi-particle excitations in the anti-bonding case as function of $k_x a$ for $k_y a = 0$ and $k_z a = 0$. Since the spin-up conduction band is partially empty, the spin-up electron conduction is possible. **(d)** A plots of QPs in the anti-bonding case as function of dimensionless momentum $k_x a$ for $k_y a = \pi/2$ and $k_z a = 0$. Since the spin-down valence band is partially empty , the spin-down hole conduction is possible. The numerical values of the parameters to be used in the calculation are t = 1 , t'/t = −0.28, t ''/t = 0.1, t '''/t = 0.06, $t_b/t$ = 0.3, $t_z/t$ = 0.1, $\frac{\alpha_0}{t} = 0.53$, $\frac{\Delta_0^{PG}}{t} = 0.01$, and $a_0 = 0.4$.

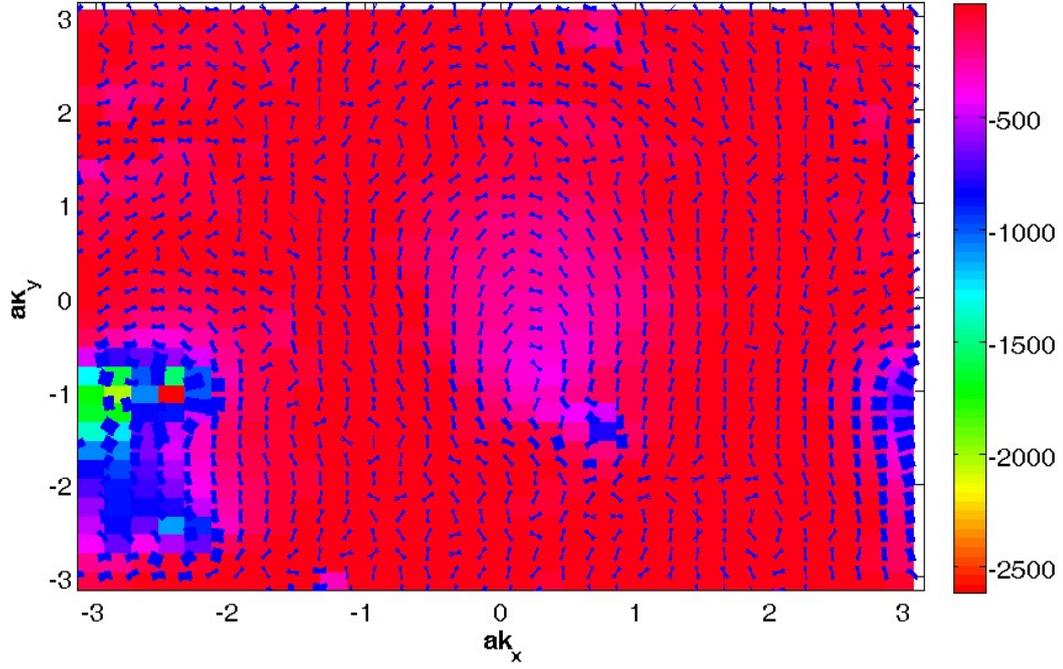

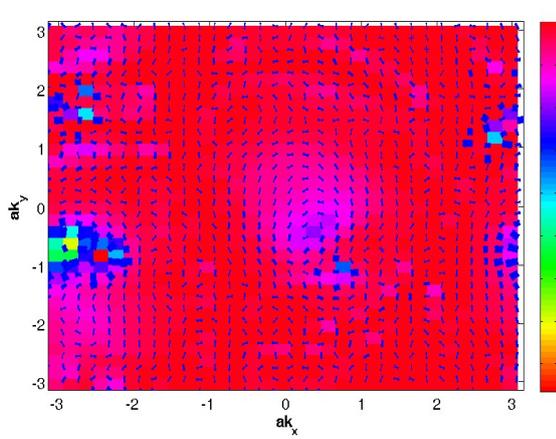

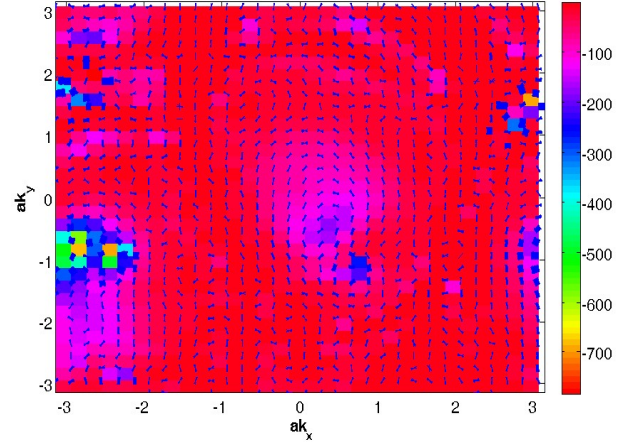

**(a)**

**(b)**          **(c)**

**Figure 4.** Contour plots of spin textures $s_z(n,\mathbf{k})$. (a) $a_0 = 0.40$ and $\alpha_0 = 0.35$. (b) $a_0 = 0.40$ and $\alpha_0 = 0.65$. (c) $a_0 = 0.60$ and $\alpha_0 = 0.65$. The numerical values of the other parameters used in the calculation are t = 1, t′/t = −0.28, t″/t = 0.1, t‴/t = 0.06, $t_b/t$ = 0.3, $t_z/t$ = 0.1, $\frac{\Delta_0^{PG}}{t} = 0.01$, μ = 0.00, $Q_1$ = 0.7742.*pi, and $Q_2$ = 0.2258.*pi. The band index 'n' stands for the spin down bands $\epsilon = -\sqrt{\frac{z_0(k)}{2}} + \varepsilon_k^U \pm \left(b_0(k) - \left(\frac{z_0(k)}{2}\right) - c_0(k)\sqrt{\frac{2}{z_0(k)}}\right)^{\frac{1}{2}}$ intersecting with Fermi energy.

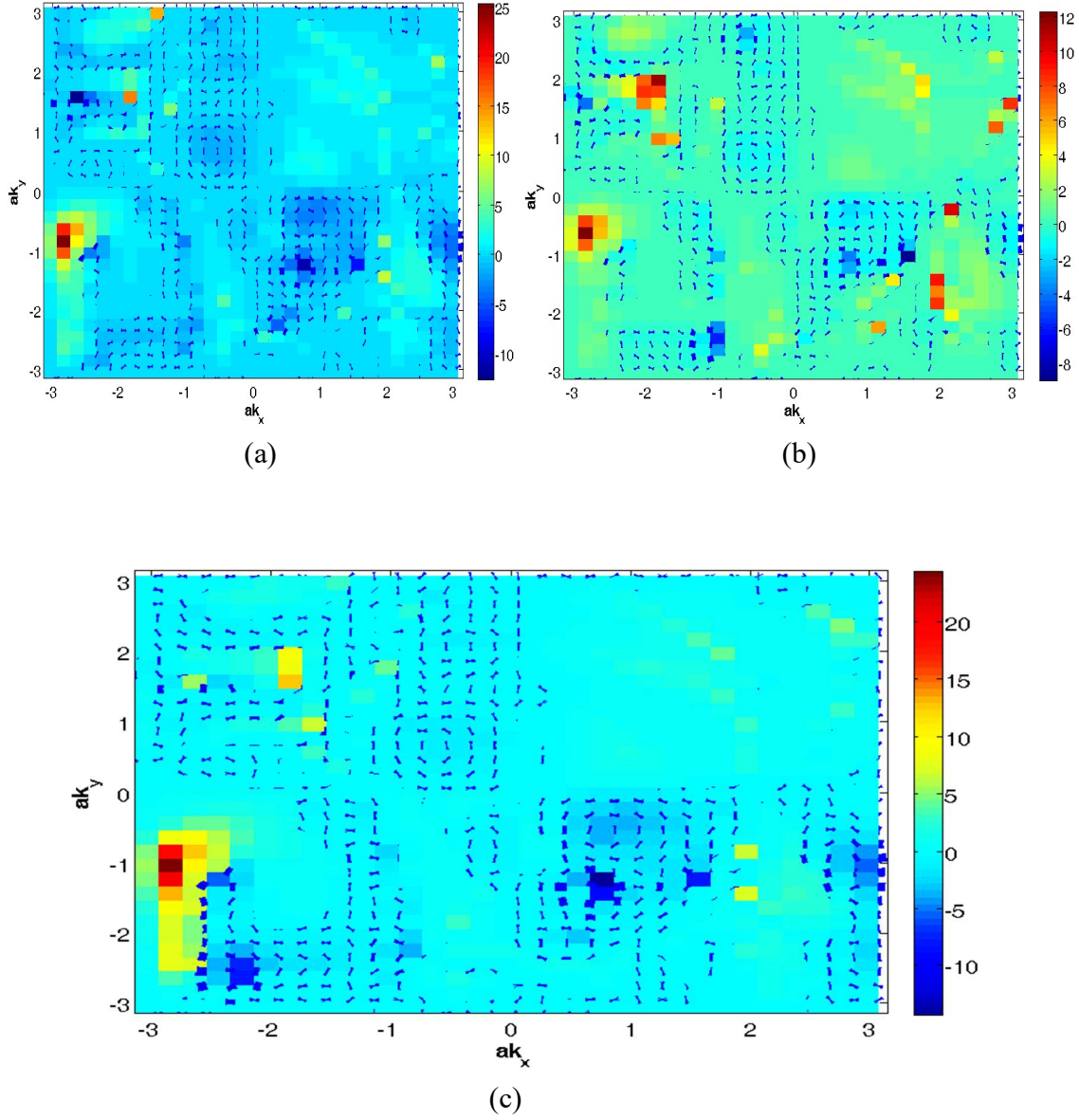

**Figure 4.** Contour plots of the expectation value ($\langle n|S_x n\rangle - \langle n|S_y|n\rangle$). (a) $a_0 = 0.40$ and $\alpha_0 = 0.50$. (b) $a_0 = 0.40$ and $\alpha_0 = 0.70$. (c) $a_0 = 0.70$ and $\alpha_0 = 0.50$. The numerical values of the other parameters used in the calculation are t = 1, t′/t = −0.28, t″/t = 0.1, t‴/t = 0.06, $t_b$/t = 0.3, $t_z$/t = 0.1, $\frac{\Delta_0^{PG}}{t} = 0.01$, μ = 0.00, $Q_1$ = 0.7742.*pi, and $Q_2$ = 0.2258.*pi. The bands $\in (\downarrow, \sigma = \pm 1, k) = -\sqrt{\frac{z_0(k)}{2}} + \varepsilon_k^U \pm \left(b_0(k) - \left(\frac{z_0(k)}{2}\right) - c_0(k)\sqrt{\frac{2}{z_0(k)}}\right)^{\frac{1}{2}}$, intersecting with Fermi energy, are chosen for the purpose of the plot.